\documentclass{article}

\PassOptionsToPackage{round}{natbib}

\usepackage[dblblindworkshop, final]{neurips_2025}

\usepackage[utf8]{inputenc} 
\usepackage[T1]{fontenc}    
\usepackage{hyperref}       
\usepackage{url}            
\usepackage{booktabs}       
\usepackage{amsfonts}       
\usepackage{nicefrac}       
\usepackage{microtype}      
\usepackage{xcolor}         

\usepackage{amssymb}
\usepackage{amsmath}
\usepackage{graphicx} 
\usepackage{float}

\workshoptitle{AI for Music}
\title{Segment-Factorized Full-Song Generation on Symbolic Piano Music}

\author{%
Ping-Yi Chen$^{1}$ \quad Chih-Pin Tan$^{2}$ \quad Yi-Hsuan Yang$^2$\\
$^1$ National Cheng Kung University \quad $^2$National Taiwan University \\
\texttt{\{a931eric,tanchihpin0517\}@gmail.com},
\texttt{yhyangtw@ntu.edu.tw}
}

\begin{document}

\maketitle

\begin{abstract}

We propose the Segmented Full-Song Model (SFS) for symbolic full-song generation. The model accepts a user-provided song structure and an optional short seed segment that anchors the main idea around which the song is developed. By factorizing a song into segments and generating each one through selective attention to related segments, the model achieves higher quality and efficiency compared to prior work. To demonstrate its suitability for human–AI interaction, we further wrap SFS into a web application that enables users to iteratively co-create music on a piano roll with customizable structures and flexible ordering.
  
\end{abstract}

\section{Introduction}
Symbolic music generation has become a prominent research topic in recent years. 
Previous studies have explored various aspects, including model architectures~\citep{hadjeres2017deepbach,huang2018music,min2023polyffusion,yuan2025diffusion},
representations~\citep{huang2020pop,compoundword}, 
controllability~\citep{wu2023musemorphose}, 
arrangement~\citep{zhao2021accomontage,tan2024picogen,tan2024picogen2}, 
and structural modeling~\citep{wu2023compose,tan2022melody,theme,wholesong}. 
Among these tasks, full-song generation remains particularly challenging, as models must not only generate long sequences efficiently but also preserve coherence across the overall song structure.

Consider how human composers typically work. 
A composer often begins by devising a theme and a high-level song structure, places the theme within the song, and then fills in the remaining sections. 
This process is only partially autoregressive: when composing a specific section, the composer usually refers to the most relevant context rather than revisiting the entire song, which would be both impractical and inefficient. 
\citet{wholesong} proposed a four-stage generation framework, which we refer to as \textit{WholeSong}, adopting a coarse-to-fine hierarchical approach with selective autoregressive conditioning to better model global song structure.
However, their approach relies on a diffusion backbone, which requires executing the entire diffusion process for each segment, resulting in an inefficient generation procedure. 
Moreover, their system encodes all context uniformly into the same conditioning space, without incorporating higher-level concepts such as themes or motifs.

In this paper, we propose \textbf{S}egmented \textbf{F}ull-\textbf{S}ong generation model (SFS), which decomposes a full song into segments using a rule-based segmentation algorithm and employs a customized Transformer to generate each segment autoregressively, conditioned only on structurally relevant preceding segments and context explicitly defined by the given structure. 
By flexibly selecting relevant segments, our model aligns with the way human composers approach songwriting: first envisioning a theme, then composing segments in a flexible order, which enables a more natural and interactive workflow in application. 
In the experiments, we show that our model outperforms the approach of WholeSong in terms of structural coherence and motif awareness, as evaluated in a user study.
We open-source our model implementation and trained weights\footnote{\url{https://github.com/eri24816/segmented-full-song-gen}}
, as well as an interactive web interface for the model.\footnote{\label{ui-url}\url{https://github.com/eri24816/co-compose}} A demo page is also provided for listening to generations from our model.\footnote{\url{https://sfs-demo.eri24816.tw}}

\section{Methodology}
\begin{figure}
    \centering
    \includegraphics[width=0.8\linewidth]{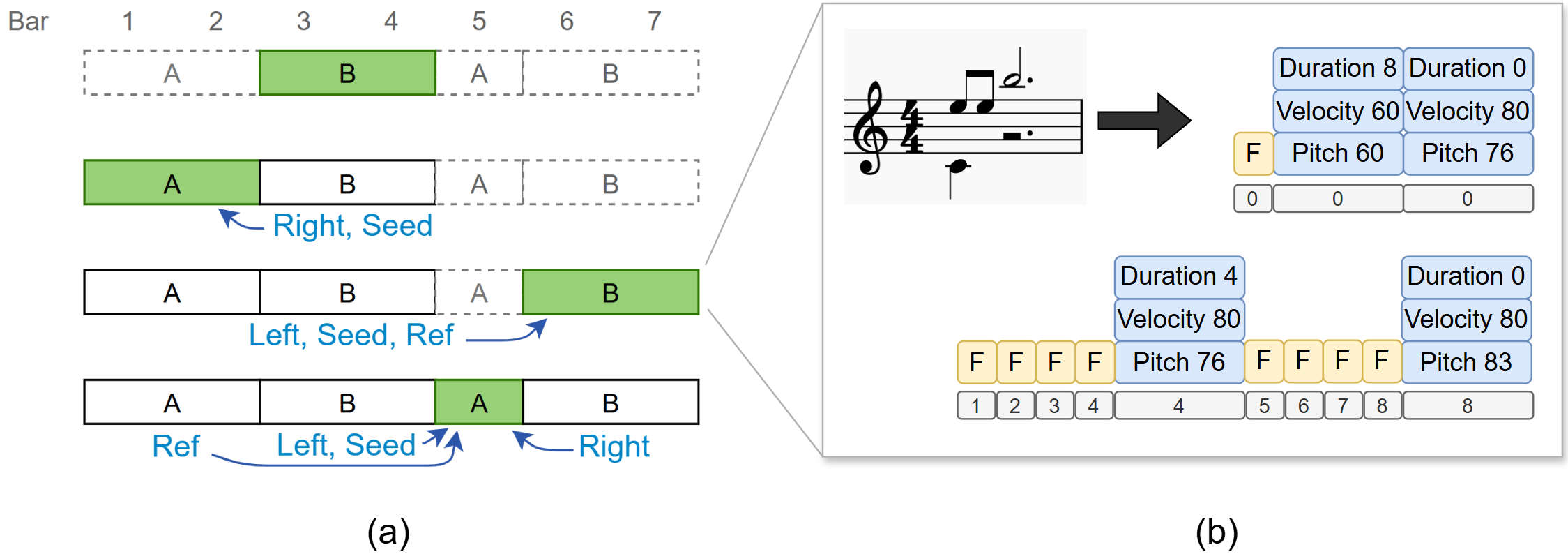}
    \caption{
    (a) Generative process with \((\hat s_1,\hat e_1)=(1,2)\), \((\hat s_2,\hat e_2)=(3,4)\), \((\hat s_3,\hat e_3)=(5,5)\), \((\hat s_4,\hat e_4)=(6,7)\), \(\hat l_{1:4}=(A,B,A,B)\), and \(o_{1:4}=(2,1,4,3)\). See Section~\ref{Segment-Factorized Full-Song Generation} for notation. (b) Example of our music language: orange refers to frame tokens, blue refers to note tokens, and gray refers to inferred positions. The \texttt{[Duration 0]} token indicates that a note’s offset is set by the next onset of the same pitch or the next bar line.
    }
    \label{fig:segments_and_tokens}
\end{figure}

\subsection{Segment-Factorized Full-Song Generation}
\label{Segment-Factorized Full-Song Generation}

SFS focuses on transforming a fixed music structure into a complete song while adhering to specified themes. 
Therefore, we assume that the music structure\footnote{\texttt{Music structure} refers to the combination of musical form and lengths of segments.} is provided by the user.
Given a song with $N$ bars $\{B_1, B_2, \dots, B_N\}$ and $M$ segments, we represent its structure as $(\hat s_{1:M}, \hat e_{1:M}, \hat l_{1:M})$, where $(\hat s_i, \hat e_i, \hat l_i)$ denote the start bar, end bar, and label of the $i^{\text{th}}$ segment, respectively, indexed from the beginning of the song.
Rather than generating music strictly in chronological order, our model is capable of generating segments in an arbitrary order.
Let $\{o_1, \dots, o_M\}$ denote the order in which the segments are generated\footnote{We train the model on all permutations, enabling adaptation to any user-specified order at inference.}. Using this order, we define the annotations $s_i = \hat s_{o_i}$, $e_i = \hat e_{o_i}$, and $l_i = \hat l_{o_i}$, so that $(s_i, e_i, l_i)$ specifies the $i^{\text{th}}$ segment to be generated.
We can then factorize the joint probability of the entire song as:
\begin{equation}
P(B_{1:N} \mid \hat s_{1:M}, \hat e_{1:M}, \hat l_{1:M}) 
= \prod_{i=1}^{M} P\big(B_{s_{i}:e_{i}} \,\big|\, B_{s_{1}:e_{1}}, \dots, B_{s_{{i-1}}:e_{{i-1}}}, s_{1:M}, e_{1:M}, l_{1:M}\big).
\end{equation}
For simplicity, we refer to the $i$th segment $B_{s_{i}:e_{i}}$ as $\mathrm{Seg}_{i}$ in the following discussion.

From our observations, musicians tend to focus on the main musical idea while selectively attending to specific contexts to ensure coherence across the song.
Inspired by this,
we define four types of essential information (referred to as the \emph{context}): $\mathrm{Left}$, $\mathrm{Right}$, $\mathrm{Seed}$, and $\mathrm{Ref}$.  
Here, $\mathrm{Left}$ and $\mathrm{Right}$ denote the nearest existing segments to the left and right of the target segment, $\mathrm{Seed}$ represents the segment carrying the main idea of the song, and $\mathrm{Ref}$ is a reference segment with the same label among the existing segments (Figure~\ref{fig:segments_and_tokens}(a)).  
The precise definition of the context is provided in Appendix~\ref{Context}.  
Instead of attending to all previously generated tokens, the model attends to these four segments at the token level, while all existing segments are encoded into a compact representation through a global vision module $G$.  
The approximated joint probability is formulated as:
\begin{multline}
P(B_{1:N} \mid s_{1:M}, e_{1:M}, l_{1:M}) \\
\approx \prod_{i=1}^M P\Big(\mathrm{Seg}_{i} \,\Big|\, \mathrm{Left}_{i}, \mathrm{Right}_{i}, \mathrm{Seed}_{i}, \mathrm{Ref}_{i}, s_{i}, e_{i}, G(\mathrm{Seg}_{1:{i-1}}, s_{1:{i-1}}, e_{1:{i-1}})\Big),
\end{multline}
where $\mathrm{Left}_{i}, \mathrm{Right}_{i}, \mathrm{Seed}_{i},$ and $\mathrm{Ref}_{i}$ are token-level references extracted from previously generated segments $\mathrm{Seg}_{1:i}$ that are most relevant to the current segment according to the given structure.  
The global vision module $G$ provides a coarse summary of all previously generated segments, giving the model awareness of the song’s overall content.






\subsection{Tokenization}


We represent music using a frame-based representation.  
The note onsets and offsets are first quantized into frames of frames of length $1/8$ beat. For each frame, a \emph{frame token} is added to the sequence, followed by \emph{note tokens} representing the notes that onset at that frame, sorted in ascending pitch order. The sequence then continues with the frame token and note tokens of the next frame.  
A note token is composed of three sub-tokens: pitch, velocity, and duration (see Figure~\ref{fig:segments_and_tokens}(b)). Positional information is encoded using model-specific embeddings (see Appendix~\ref{pe} for details).




\subsection{Implementation}
\label{Model Implementation}

\begin{figure}
    \centering
    \includegraphics[width=1\linewidth]{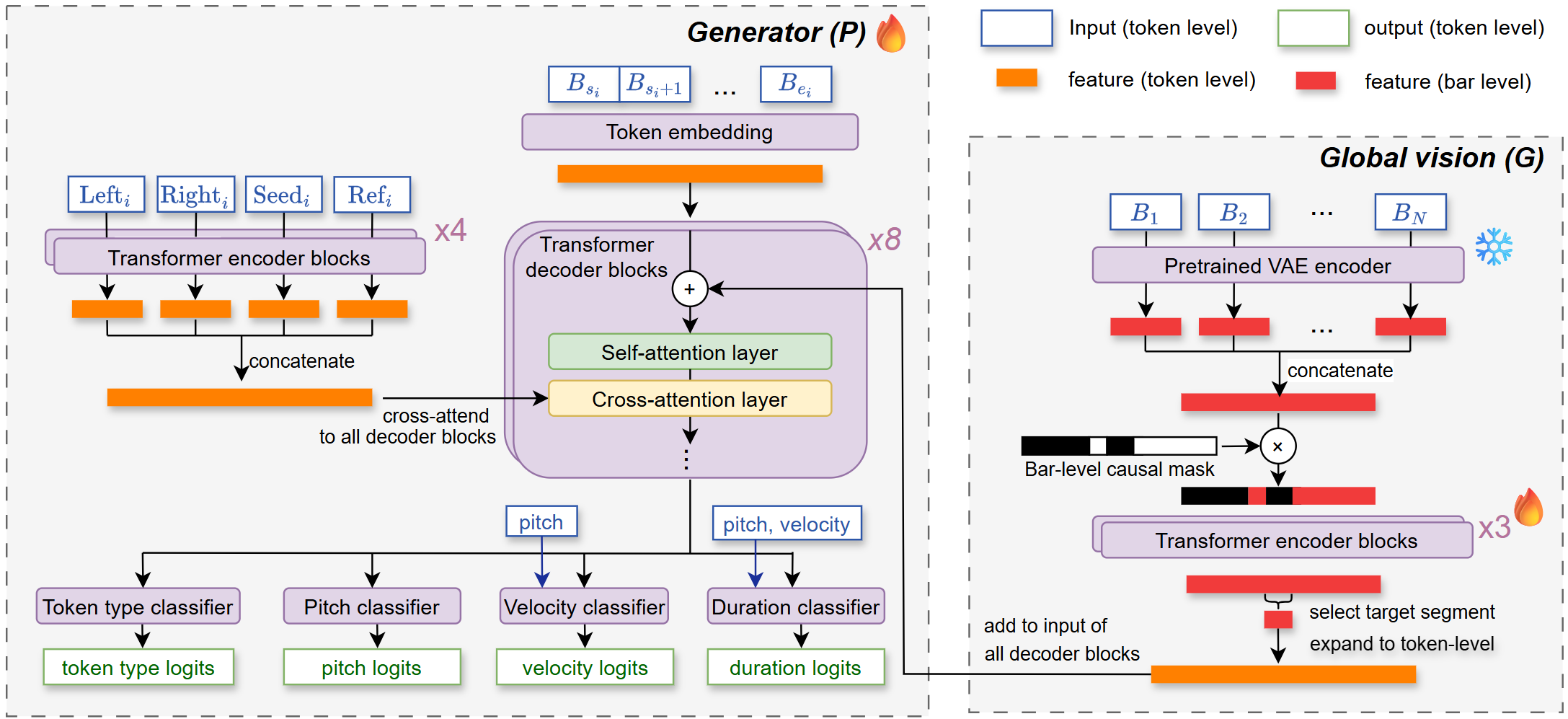}
    \caption{Model architecture of the Segmented Full-Song Model. At the output heads of the Generator (bottom middle), pitch, velocity, and duration are generated sequentially due to their dependencies. During training, the velocity classifier receives the ground-truth pitch, and the duration classifier receives the ground-truth pitch and velocity. During inference, they instead receive sampled values.}

    \label{fig:model implementation}
\end{figure}
The model consists of two components: the global vision encoder $G$ and the song generator $P$ (see Figure~\ref{fig:model implementation}).  
The global vision encoder $G$ employs a pretrained VAE encoder to embed each bar in $\mathrm{Seg}_{1:{i-1}}$, converting tokens into bar-level embeddings. 
The generator $P$ is a Transformer conditioned on two sources: (i) the global vision output, incorporated into the decoder via in-attention~\citep{wu2023musemorphose}, and (ii) the context $(\mathrm{Left}_i, \mathrm{Right}_i, \mathrm{Seed}_i, \mathrm{Ref}_i)$, provided through the generator’s encoder. 
To represent positional information in both $G$ and $P$, we apply positional encodings specifically designed for our model, with details provided in Appendix~\ref{pe}.


\section{Evaluation and Discussion}
\begin{table}[]
  \caption{Comparision between models}
  \label{sample-table}
  \centering
  \begin{tabular}{lllllll}
    \toprule
    & & \multicolumn{3}{c}{SI} & \multicolumn{2}{c}{User Study}               \\
    \cmidrule(r){3-5} \cmidrule(r){6-7} 
    \textbf{Model}     & \textbf{Inference Speed}     & $\textbf{SI}_{2-8}$  & $\textbf{SI}_{8-16}$& $\textbf{SI}_{16+}$& \textbf{O} & \textbf{A}\\
    \midrule
    SFS (Ours) & 2.03 beat/sec.  &0.3286 & \textbf{0.2264}& \textbf{0.1109}& 3.14 & \textbf{3.59}  \\
    WholeSong     &  0.197 beat/sec. & 0.3234 &0.2262&0.0860  & 3.02& 3.16 \\
    
    Flat     & 5.68 beat/sec. & \textbf{0.3426}& 0.1990& 0.0409 &\textbf{3.36} &2.34 \\
    \midrule
    Datset     & -  & 0.4398 & 0.3827 & 0.3300 & 4.00 &4.07 \\
    \bottomrule
    \\
  \end{tabular}
\end{table}

We compare our model against two baselines: a \textit{flat} model and WholeSong. The flat model is a GPT-like Transformer trained on 8-bar fragments, which can be regarded as an ablation of our model without structural conditioning. 
For a fair comparison, we only use the last 3 levels of Wang's model for inference because the form is a given condition. Also, we modify the inference code of WholeSong to support seed conditioning by fixing the given seed segment throughout the diffusion process across all three levels of generation.
All models are trained on our in-house dataset, consisting of 32,090 pop piano performances for training and 3,615 for testing.

For objective evaluation, we sample 45 songs from the test set. Using their structural specifications, we condition both our model and WholeSong to generate corresponding songs. The flat model instead simply generates unconditional pieces of the same length. We then compute the average Structureness Indicator (SI) \citep{jazz} at short (2–8 bars), mid (8–16 bars), and long (16+ bars) ranges, which reflects the ability of a model to maintain structural consistency.

For subjective evaluation, we generate 16 samples with our model and WholeSong, conditioned on seed segments and structures from the test set. The flat model again produces unconditioned sequences. We then conduct a user study in which each participant listens to the seed, followed by generations from the models and the original song. Participants rate each piece on a 1–5 scale along two aspects: \textbf{O}verall quality (O) and \textbf{A}dherence to seed (A). We collect responses from 44 participants and report the mean scores.

Results of both evaluations are shown in Table~\ref{sample-table}. Compared to WholeSong, our model achieves stronger adherence to the seed and slightly higher scores in both overall quality and structureness. However, a substantial gap remains relative to real data, highlighting the need for further improvement in full-song generation. We suspect that the higher overall quality reported for the flat model arises from its fluency, which is often preferred by users, as it always generates in a forward direction.

In terms of efficiency, our model achieves real-time generation at an average of 2.03 beat/second
, about 10$\times$ faster than WholeSong, despite operating at twice the temporal resolution (1/8 beat vs. 1/4 beat). This real-time capability enables streaming output to a user interface during generation, improving user experience in applications such as interactive composition and live performance.

Despite the segment-level correspondence, we notice that the generated music sometimes lack smooth phrase-level transitions and progression across the full song. The segments may appear thematically consistent yet loosely connected, without the natural buildup and flow that typically define sections such as the introduction, verse, chorus, and outro. This suggests the need for a higher-level planning mechanism to guide how each phrase develops within the overall song form in future works.

\section{Web Interface}

We build a web interface for our model to showcase human–AI co-composition.\footref{ui-url} It provides (i) a structure editor for defining song structure and seed, and (ii) a piano-roll editor where users can edit notes or let the model generate selected ranges, enabling iterative, arbitrary-order song completion. The application also serves as a general interface for full-song music generation via an abstract Python API that can connect to different models. Usage instructions are in Appendix~\ref{application}.

\section{Conclusion}

We proposed the Segmented Full-Song Model (SFS) and demonstrated its capability in generating complete songs. Experiments show that SFS achieves stronger adherence to seeds, improved structural consistency, and higher subjective quality compared to prior work, while being an order of magnitude faster, enabling real-time streaming. We also highlighted its potential for human–AI co-creation through a web-based composition interface.
We look forward to further explorations of model designs that align more closely with human creative workflows and open new possibilities for human–AI interaction in music-making.

\newpage
\bibliography{bib}
\newpage


\appendix

\section{Definition of the Context}
\label{Context}

To provide contextual information for generating a target segment \(\mathrm{Seg}_{i}\), the context is selected from segments that have been generated, \(\mathrm{Seg}_{{1:i-1}}\), with the following rule:

\begin{itemize}

\item  Left$_{i}$: the nearest existing segment to the left of $\mathrm{Seg}_{i}$, helps generate a smooth transition from that segment.
  \[
    \mathrm{Left}_i = \mathrm{Seg}_n \quad \text{where} \quad n = \arg\max_{j<i} \{e_j \mid e_j \le s_i\}
  \]
  (or $\varnothing$ if no such $j$ exists)

\item  Right$_{i}$: the nearest existing segment to the right of $\mathrm{Seg}_{i}$, helps generate a smooth transition to that segment.
  \[
  \mathrm{Right}_i = \mathrm{Seg}_n \quad \text{where} \quad n = \arg\min_{j<i} \{s_j \mid s_j \ge e_i\}
  \]
  (or $\varnothing$ if no such $j$ exists)

\item Seed$_i$: The seed segment carries the main idea of the song (motifs and overall style) to all other segments. It is defined as the first segment of the song, based on the assumption that the main idea is the first thing to be written down when composing music.
  \[
  \mathrm{Seed}_{i} =
  \begin{cases}
    \varnothing & \text{if } i = 1 \\
    \mathrm{Seg}_{1} & \text{if } i > 1
  \end{cases}
  \]

\item Ref$_i$: A reference segment with the same label occurring earlier. The earliest segment is chosen if multiple segments are eligible:
  \[
    \mathrm{Ref}_i = \mathrm{Seg}_n \quad \text{where } n = \min_j \{ j \mid l_j = l_i, j < i \}
  \]
(or $\varnothing$ if no such $j$ exists)

\end{itemize}

In practice, to limit computational cost, each of the four context types is truncated to a maximum length of 8 bars. For Left, we retain the right-most 8 bars, while for Right, Seed, and Ref, we retain the left-most 8 bars.









\section{Experiment Setting}
The dataset consists of performances of piano covers of pop music from YouTube, ranging from 4 to 200 bars.  On average, each song contains 4,394 tokens, or 54.6 tokens per bar. It includes 32,090 songs for training and 3,615 songs for testing. All songs are transported to C major or A minor key. The structural labels of the pieces are assigned automatically using our segmentation algorithm described in Appendix~\ref{Details about the segmentation algorithm}.

The procedure to construct one training sample is:
\begin{enumerate}
    \item Sample a song in the dataset.
    \item Obtain its structure $(\hat s_{1:M},\hat e_{1:M},\hat l_{1:M})$ using the segmentation algorithm.
    \item Identify the label that appears on the most bars, in which we assume the theme is located. Among the segments with that label, select the one that starts closest to the song's middle, and assign it to $o_1$.
    \item Assign a segment to $o_2$ with similar way we assign $o_1$, but select the second-most frequent label.
    \item Fill $o_{3:M}$ with the remaining segments in a random order to construct an arbitrary sequence $o$ as specified by the user.
    \item Randomly pick one segment to be the target segment. If its length is greater than 8 bars, randomly sample an 8-bar fragment inside it, so it can fit our model's receptive field.
    \item Obtain the context segments with the definition described in Appendix~\ref{Context}.

\end{enumerate}

We train the model for 2 million steps over 127 hours on a single RTX~4090 GPU. The batch size is 12. Optimization is performed with Adam, minimizing negative log-likelihood (NLL) \textit{summed} over tokens rather than averaged. The learning rate decays exponentially from $1\times10^{-4}$ to $5\times10^{-6}$ through the training.

\section{Details about the segmentation algorithm}

\label{Details about the segmentation algorithm}
\subsection{Similarity Metric}

The segmentation algorithm is based on our similarity metric between two bars of music $s(\cdot,\cdot)$. It is constructed as follows.

First, given two bars of music $a$ and $b$, each considered as a set of notes, we define the \textit{note overlap score} $f(a,b)$ as the maximum number of one-to-one matches between notes in $a$ and $b$, divided by $\max(|a|,|b|)$. A valid match requires that the notes share the same pitch and that their onset times differ by at most one frame (i.e., $1/8$ beat).

To increase the weight of the skyline in the note overlap score, we define the skyline of a bar
\[
\mathrm{SKY}(a) = \Bigl\{ n \in a \,\Big|\,
\nexists m \in (a-\{n\}) \;\text{such that}\;
\frac{m.\mathrm{pitch} - n.\mathrm{pitch}}{|m.\mathrm{onset}-n.\mathrm{onset}|}
> \tfrac{1\ \text{octave}}{1\ \text{beat}}
\Bigr\}
\]
and revise the overlap score to jointly consider all notes and skylines:
\[
\tilde{f}(a,b) = 0.5\,f(a,b) + 0.5\,f(\mathrm{SKY}(a),\mathrm{SKY}(b)).
\]

Finally, the similarity between two bars is defined as
\[
s(a,b) = \max\bigl(\tilde{f}(a,b),\, \tilde{f}(a,b^{\textit{8va}}),\, \tilde{f}(a,b^{\textit{8vb}})\bigr),
\]
where $b^{\textit{8va}}$ and $b^{\textit{8vb}}$ denote transpositions of $b$ one octave up and down, respectively.

\subsection{Segmentation Algorithm}
Given a song divided into $N$ bars $B_1, \dots, B_N$, the segmentation proceeds as follows:

\begin{enumerate}
    \item \textbf{Similarity matrix.}  
    Compute the bar-wise similarity matrix
    \[
    \mathbf{S}_{i,j} = s(B_i, B_j),
    \]
    where $s(\cdot,\cdot)$ is the similarity between two bars.

    \item \textbf{Adjacency regularization.}  
    This step encourages consecutive label assignment. First, construct a banded adjacency matrix $\mathbf{A}$ with ones on the main diagonal and the first diagonals above and below it.  
    Then form the adjusted similarity matrix
    \[
    \mathbf{M} = (1-\alpha)\max(0.3,\mathbf{S}) + \alpha \mathbf{A}.
    \]

    \item \textbf{Spectral embedding.}  
    Compute the unnormalized graph Laplacian $\mathbf{L} = \mathbf{D} - \mathbf{M}$,  
    where $\mathbf{D}$ is the diagonal degree matrix.  
    Let $\lambda_1 \le \cdots \le \lambda_N$ be the eigenvalues of $\mathbf{L}$, and choose
    \[
    k = \arg\max_{2 \le j \le K_{\max}} (\lambda_{j} - \lambda_{j-1}),
    \]
    i.e., the index of the largest eigen-gap up to $K_{\max}$.

    \item \textbf{Clustering.}  
    Take the first $k$ eigenvectors of $\mathbf{L}$, normalize them, and run $k$-means clustering to obtain bar labels $\ell_1, \dots, \ell_N$.

    \item \textbf{Segmentation.}  
    Identify split points at positions where $\ell_i \neq \ell_{i-1}$. Using these split points, the start bar, end bar, and label of each segment in the song are determined.

\end{enumerate}

In practice, we adopt
\[
\alpha = 0.7, 
\quad 
K_{\max} = \min\!\Big(6, \Big\lfloor \tfrac{N}{8} \Big\rfloor\Big),
\]
which we find to work well on our dataset.

We spot two limitations of the algorithm. First, it does not provide semantic labels (verse, chorus, etc.). Also, it can't identify consecutive repeated segments. Instead of reporting two segments with the same label, it reports one big segment.
\begin{figure}[H]
    \centering
    \includegraphics[width=1\linewidth]{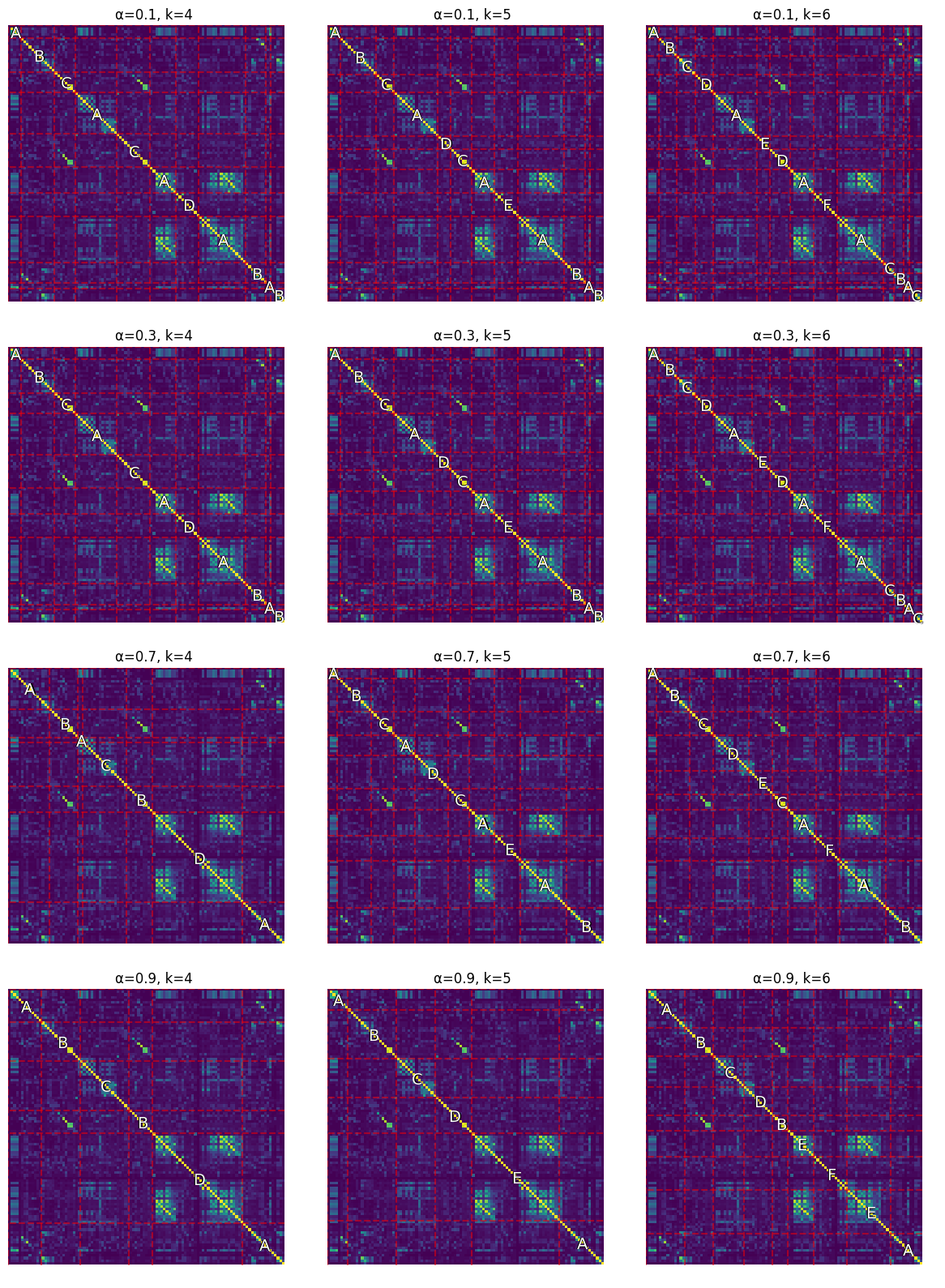}
    \caption{The segmentation result of song \textit{
Hikaru Nara - Your Lie in April OP1 [Piano]}(\url{https://www.youtube.com/watch?v=zsVAbS8xmaU}) using our algorithm with different settings of $\alpha$ and $k$. The setting we actually use for this song $\alpha=0.7$ and $k=6$. The heatmap shows the similarity matrix of the song, where purple to yellow indicates 0 to 1.}
    \label{fig:placeholder}
\end{figure}



\section{Positional Encodings Used in Our Model}

\label{pe}
The positional information is not directly embedded into frame tokens but instead provided through positional embeddings. When predicting a token, the position assigned to the query of each attention block always corresponds to the token’s position in music, measured in frames.  

Three types of positional encodings are used in our model: Start-End positional encoding (Start-End PE), Sub-beat positional encoding (Sub-beat PE), and RoPE. We use a concatenation of Start-End PE relative to song, Start-End PE relative to segment, and Sub-beat PE and add it to every token-level input to provide complete information about the local and the global position. RoPE is used in every self-attention and cross-attention layer.

Start–End positional encoding provides the model with information about a token’s relative position within a segment or song. It is constructed by concatenating two standard sinusoidal encodings: one representing the distance from the start and the other representing the distance from the end.

\begin{figure}[H]
    \centering
    \includegraphics[width=0.3\linewidth]{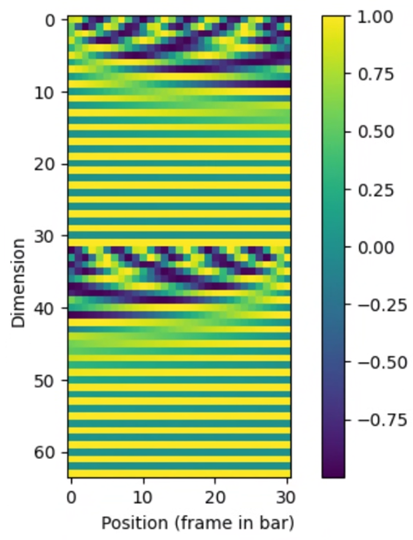}
    \caption{Start–End positional encoding}
    \label{fig:placeholder}
\end{figure}

Sub-beat positional encoding indicates the frame index inside a bar. It is a concatenation of a one-hot vector and a binary vector.
\begin{figure}[H]
    \centering
    \includegraphics[width=0.4\linewidth]{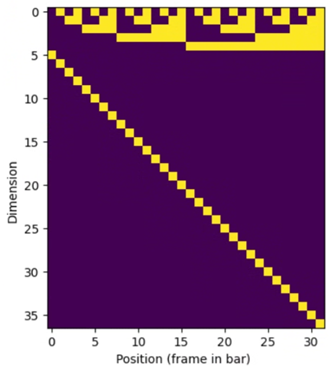}
    \caption{Sub-beat positional encoding. Purple indicates 0, and yellow indicates 1.}
    \label{fig:placeholder}
\end{figure}

Notably, all positional embedding in our model uses time frames (instead of tokens' positions in sequence) as position indices, which provides a more natural representation. 

\section{Details about the User Study}
To construct the listening materials for the user study, we sample 16 songs from the test set whose seed segments consist of 8 bars. For each song, we use the seed and structure to condition our model and WholeSong. For the flat model, we generate songs of the same length without additional conditioning.

In the user study, each participant evaluates four instances of musical pieces associated with the same seed and structure in a shuffled order. These four instances are: generation from our model, generation from WholeSong, generation from the flat model, and the real sample from the dataset. In the study, we refer to these as \textit{generations}.

 For each generation, the participant first listens to the seed, then to the generation, provided with the following description:

\begin{itemize}
    \item \textbf{Seed:} a given short musical idea.
    \item \textbf{Generation:} a complete piece of music generated based on the seed.
\end{itemize}

Participants then rate each generation on a 5-point Likert scale (1 = lowest, 5 = highest) according to the following criteria:

\begin{enumerate}
    \item \textbf{Adherence to Seed:} \\
    How much does the generated piece retain the seed’s musical idea? How similar is the overall style or mood of the generated piece compared to the seed?

    \item \textbf{Structureness:} \\
     How good is the generated piece's structure as a complete composition? Consider, for example, does the piece have a clear form (e.g. intro, verse, chorus, outro etc.) and have a reasonable emotional development?

    \item \textbf{Overall Quality:} \\
    How good is the overall quality of the generated piece? How good does the generated piece sound to you?
\end{enumerate}

We collect 44 responses from 21 amateur participants (with no or fewer than 3 years of music-related training), 19 experienced participants (3 or more years of training), and 4 professionals, and we report the mean scores.

\section{Instructions for Using Our Web Interface}
\label{application}
\begin{figure}
    \centering
    \includegraphics[width=1\linewidth]{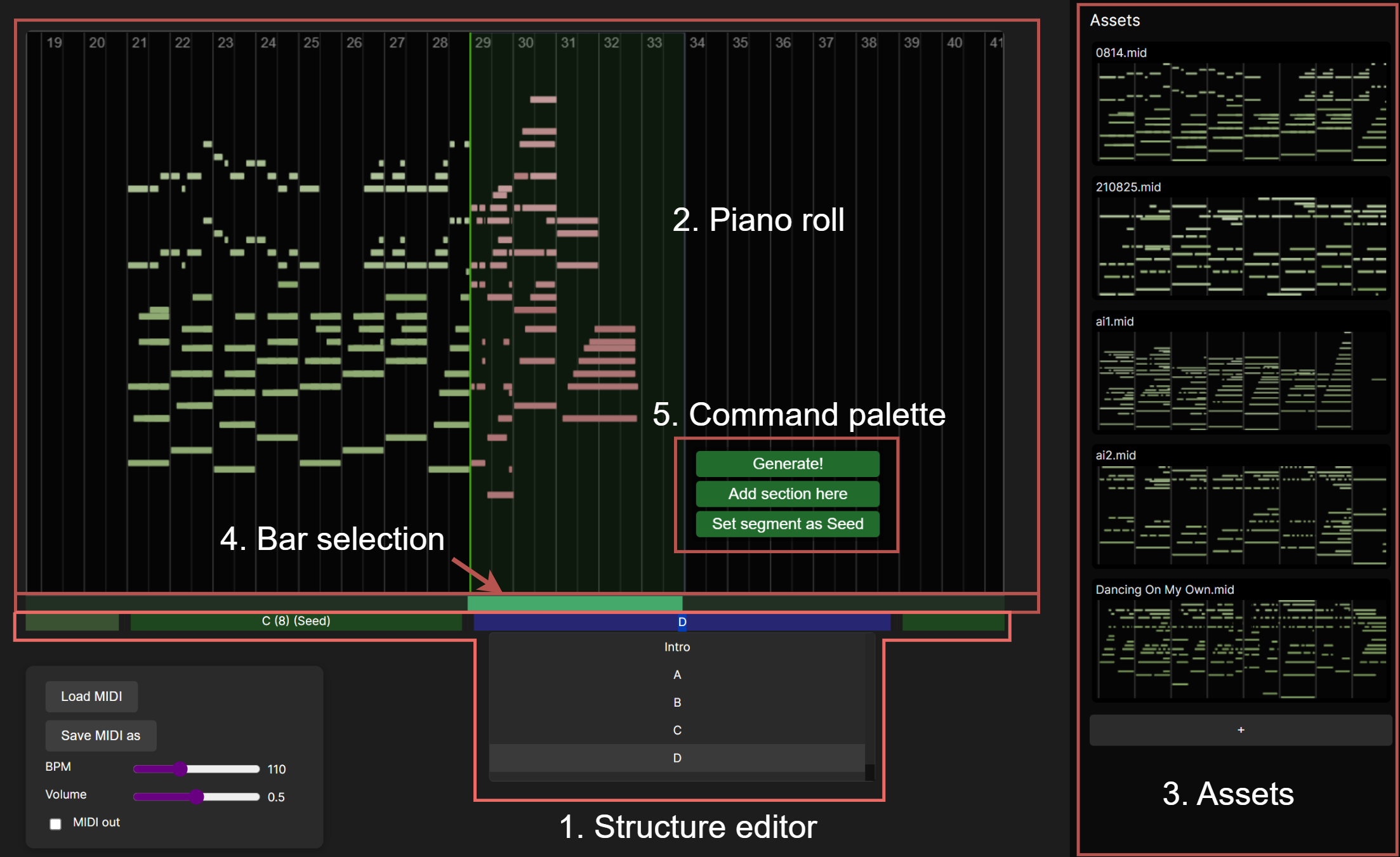}
    \caption{User interface of our web Interface.}
    \label{fig:placeholder}
\end{figure}

The interface consists of the following components:

\begin{enumerate}
    \item \textbf{Structure editor.} Lets users define the song structure, which serves as a condition for generation. The structure can be specified beforehand or adjusted during composition. Drag the left or right edge of a segment to resize it, or click a segment to assign its label.  
    \item \textbf{Piano roll.} The workspace where users and AI collaborate on music. Users can use the space bar to toggle play, click on empty space to create a note, drag to move a note, and right-click to delete a note. Use the scroll wheel to pan, and hold \texttt{Control} while scrolling to zoom.  
    \item \textbf{Assets.} Provides several 8-bar MIDI assets that users can drag into the piano roll as starting material or to combine with an existing composition. Users can also import additional assets from their own disk. Click on an asset to preview it.
    \item \textbf{Bar selection.} Enables quick selection of one or multiple bars, which can then be used with the command palette.  
    \item \textbf{Command palette.} Click \textit{Generate!} to let the AI generate content for the selected range, or use other buttons to perform different operations.  

\end{enumerate}

To play with it, here’s a workflow we recommend:

\begin{enumerate}
   
\item First, specify the desired structure of the song. Using the default structure is also acceptable.
\item Next, set the seed segment. You can do this by dragging in an asset, composing it yourself in the built-in or an external editor, or letting the AI generate it. Because the application is not yet ideal for detailed note editing, we suggest composing in another program and importing the MIDI as an asset.
\item Finally, generate the remaining segments in any order, one or several bars at a time. Incorporate human composition when you have a clear idea or wish to refine the AI’s output. As our model does not allow fine-grained control, generating an entire segment at once may yield overly random results. In such cases, shorten the generation range and apply rejection sampling guided by human evaluation.

\end{enumerate}

Note that users can listen to the AI’s generation in real time by playing the music immediately after clicking the \textit{Generate!} button, as the output streams to the client. For this feature to work reliably, the BPM should be kept below about 120 (or occasionally 100), based on our tests on an RTX 4090. This limit may vary depending on hardware and runtime conditions.


\end{document}